\documentclass[multphys,vecphys]{svmult}

\usepackage{makeidx}     
\usepackage{graphicx}    
\usepackage{multicol}    

\def\ltsimeq{\,\raise 0.3 ex\hbox{$ < $}\kern -0.8 em \lower
0.7 ex\hbox{$\sim$}\,}

\makeindex             

\begin{document}

\title*{The quasar Q0957+561: Lensed CO emission from a disk at z$\sim$1.4?}
\author{M. Krips\inst{1} \and
R. Neri\inst{2} \and A. Eckart\inst{1} \and J.
Mart\'in-Pintado\inst{3} \and P. Planesas\inst{3} \and L.
Colina\inst{4}}
\authorrunning{M. Krips}
\institute{I. Phys. Inst. der Universit\"at zu K\"oln Z\"ulicher
Str. 77, 50937 K\"oln, Germany \texttt{krips@ph1.uni-koeln.de}
\and IRAM, 300, rue de la Piscine, Domaine Univ.,F-38406 Saint
Martin d'H\`eres \and Observatorio Astron\'omico Nacional (IGN),
Alcal\'a de Henares, Spain \and Instituto de Estructura de la
Materia (CSIC), Madrid, Spain}
%
%
\maketitle

In recent years large efforts have been made to detect molecular
gas towards high redshifted objects. Up to now the literature
reports on only two cases of CO-detection in quasars at a redshift
between 1 and 2 -- Q0957+561, a gravitationally lensed system at
$z=1.41$ (Planesas et al.\ 1999) , and HR10 at $z=1.44$ (Andreani
et al.\ 2000). According to Planesas et al.\ (1999),
$^{12}$CO(2$\rightarrow$1) emission was detected towards both the
lensed images of Q0957+561 with the IRAM Plateau de Bure
Interferometer (PdBI; Fig.\ 1).  In contrast to the optical
spectra of the two images which support the idea that they are
images of one and the same object, the CO-spectra were
surprisingly different: the southern image (named CO-B) shows a
single blueshifted line whereas a double-peaked line profile with
a blue- and a redshifted part appears towards the northern image
(CO-A). Based on the observations and on simulations with a
gravitational lens program, we are tempted to argue that the line
profile traces the presence of molecular gas of a disk in the host
galaxy around the quasar. We have now new observations with the
PdBI providing the necessary sensitivity to corroborate our
disk model. \\

\section{Introduction}
\label{intro}
Since the discovery of Q0957+561, the first confirmed
gravitationally lensed quasar (Walsh et al.\ 1979) at a redshift
of $z=1.4$, several models have been developed to understand the
lensing potential of the intervening galaxies: a giant elliptical
galaxy (G1) at a redshift of $z = 0.36$ with a surrounding cluster
at $z = 0.355$ and probably another group of background galaxies
at $z = 0.5$ (e.g. Angonin-Willaime et al.\ 1994). Planesas et
al.\ (1999, in the following P99) have recently achieved in
observing the CO(2-1) line in Q0957+561, but lacked sensitivity
and a detailed lensing model to interpret the origin of the
molecular emission. To confirm and substantiate this line profile,
we have carried out new observations with the IRAM interferometer
making Q0957+561 potentially very valuable for understanding the
evolution of quasars at redshifts $1<z<2$. To further improve on
P99's interpretation, we have developed a numerical code
incorporating existing lensing models of Q0957+561.

\section{Observations}
\label{obs}
CO(2-1) and (5-4) observations of Q0957+561 were done
simultaneously in 1998 and again in 2003 with the IRAM
interferometer.
\\
{\it Continuum emission:} In the 3.1mm radio continuum, two lensed
images of the quasar, labeled A and B (pointlike), and a radio jet
C (extended) appear at this wavelength (Fig.\ 1), in agreement
with VLA observations (Harvanek et al.\ 1997). The positions
coincide with the optical ones within the errors. There is no
evidence for variability above 10\% between May 1998 and April
2003 in all three components. The lack of any clear detection at
1.3mm could also be a sign for resolving out flux in the smaller
beam.
\\
{\it Line emission:} After subtracting the continuum the
integrated CO(2-1) map shows two lensed images labeled CO$-$A and
CO$-$B, as in P99. The emission centroids of the two images are
separated by $\sim1''$ more than in the optical/radio. The two CO
line components show different spectral profiles. A double-peaked
profile is visible towards the northern image CO$-$A whereas only
a single (blueshifted) velocity component is detected towards
CO$-$B. No line emission was detected towards CO(5-4).

\section{Modelling Q0957+561}
We developed a numerical code based on the standard gravitational
lens equation to explain the absence of the double-peaked line
profile towards CO$-$B in Q0957+561. The code has been applied on
6 different models of Q0957+561, all based on previous work by
Barkana et al.\ (1999). They can be mainly divided into 2 main
groups: one set of models is based on a King profiles, as proposed
by Falco et al.\ (1985$-$ models: {\footnotesize FGS, FGSE,
FGSE+CL}), the second set on a softened power-law distribution, as
suggested by Grogin et al.\ (1996$-$models: {\footnotesize SPLS,
SPEMD, SPEMD+CL}). Each of these two sets can be subdivided into
three subgroups assuming either elliptical or spherical profiles
and/or a shear term or a Single Isothermal Sphere for the
surrounding cluster. Besides the sizes of the lensed images, the
positions, continuum and the blueshifted line intensity ratios,
also the optical time delay (Kundic et al.\ 1997) has been taken into
account to distinguish among the models. For simplicity, the
respective components were simulated as slightly extended
gaussians. The {\footnotesize SPEMD+CL} and {\footnotesize
FGSE+CL} models reproduce the observed constraints with the lowest
$\chi_r^2$ ($\leq3$) showing that the contribution of the cluster
is important: The models that best explain these observations all
require a {\footnotesize SIS}
cluster.\\
{\bf\it Line emission:} Let us first forget about the redshifted
CO-A component and concentrate on the blueshifted line. A compact
region centered on the nucleus can be directly ruled out by the
small line ratio B-blue/A-blue of ~0.4 since this would be in
contradiction with the measured 3mm continuum ratio A/B of 1.5
originating in a compact region. An extended region of the
blueshifted line indeed reproduces the derived small line ratio of
0.4 but it would have a particular shape (very extended in
direction of the nuclear jet ($\sim5"$) but very thin
perpendicular to it). Also a non-nuclear position of the
blueshifted emission region with a rather compact shape and closer
to the inner tangential caustic originates in a small line ratio.

The difference in the line profiles observed towards CO$-$A and
CO$-$B can be explained by the location of the redshifted gas
component relative to the lens caustic. In the best-fit simulation
for the {\footnotesize SPEMD+CL} model the two caustics are so
close together that the blueshifted component of Q0957+561 is
still located between them whereas the redshifted line must
originate above the northern outer caustic. This calculation used
two separated components of the red- and blueshifted line but we
can also start with an extended region centered on the nucleus
where the north-eastern part corresponds to the redshifted
component and the south-western part to blueshifted one. The
morphology that reproduces the derived line ratio
(A-blue+A-red)/B-Blue$\sim 1.1$ agrees well with the model of the
host galaxy used by Keeton et al. for the Barkana FGSE+CL model
being quite close to our best-fit FGSE+CL model.

\section{Discussion and Conclusions}

What can we conclude about the origin of the redshifted CO(2-1)
velocity component? Mainly two groups of hypotheses are possible:
it either originates from an independent system (companion etc.)
or traces the presence of molecular gas in the rotating disk of
the host galaxy in Q0957+561. We have reasons to discard the first
class of hypotheses. One of the strongest arguments is that the
host galaxy detected in HST observations by Keeton et al. (2000)
is oriented in the direction of the two line components in CO$-$A
and is extended on the same scale. Also the integrated line ratio
of (CO-A-red+CO-A-blue)/CO-B $\approx$1.0 is consistent with
Keeton et al. (2000). Furthermore, the simulation of the two
velocity profiles together supports also an extended disk similar
in shape to the used host galaxy models of Keeton et al.\ 2000
whereas the model for a single blueshifted component would either
suggests a very thin but very elongated disk not at all similar to
the host galaxy detected in the optical or a quite compact CO
region offset from the nucleus. The redshift of this region would
then also be significantly different from the one of the host
galaxy and the quasar. Thus, the latter two approaches seem to be
quite unlikely compared to Keeton et al.'s (2000) results for the
host galaxy. Furthermore, the double-peaked line profile at CO$-$A
appears to be symmetric and is centered within the errors at
$z_0=1.4141$. Based on these arguments, we favour the second
hypothesis: the presence of an important reservoir of molecular
gas in a disk of the host galaxy surrounding Q0957+561. Based on
the lower magnification from the {\footnotesize FGSE+CL} model, we
have estimated an upper limit to the molecular gas mass for each
integrated blue and redshifted velocity profile in CO$-$A and
CO$-$B. Assuming a standard $M_{gas}$ to $L'_{\rm{CO(1-0)}}$ ratio
($\simeq 5\,M_\odot$(K\,km\,s$^{-1}$pc$ ^2)^{-1}\,$, Downes et al.\
1993) and $R_{21}=$CO(2-1)/CO(1-0)$\ltsimeq 1$, we find $M_{\rm
gas}=2.3\times10^{10}\,M_\odot$ for both the blue and red velocity
profiles. The low upper limit on $R_{54} \simeq 1$, the velocity
averaged line ratio, characterizes low excitation conditions, and
thus agrees with global CO emission from a disk. The agreement
between molecular gas masses obtained with the integrated CO
luminosities and individual magnification factors, each tracing
line emission from half of the quasar host, is a further support
for the rotating disk hypothesis.

%
%
%
\begin{figure}
\centering
\vspace*{-0.2cm}\rotatebox{-90}{\includegraphics[height=9.3cm]{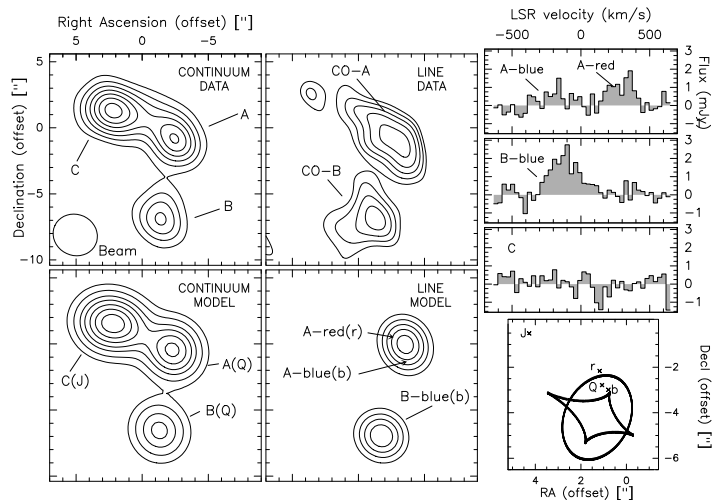}}
%
%
\caption{Upper panel: observed continuum (left) \& line emission
(middle) plus spectrum (right); lower panel: simulated data with
the lens plane (right).}
\label{fig:1}       
\end{figure}

%
%



\printindex
\end{document}